# Large coverage fluctuations in Google Scholar: a case study

Alberto Martín-Martín[1] and Emilio Delgado López-Cózar[2]

[1] albertomartin@ugr.es
Facultad de Comunicación y Documentación, Universidad de Granada (Spain)

[2] edelgado@ugr.es
Facultad de Comunicación y Documentación, Universidad de Granada (Spain)

**Abstract**
Unlike other academic bibliographic databases, Google Scholar intentionally operates in a way that does not maintain coverage stability: documents that stop being available to Google Scholar's crawlers are removed from the system. This can also affect Google Scholar's citation graph (citation counts can decrease). Furthermore, because Google Scholar is not transparent about its coverage, the only way to directly observe coverage loss is through regular monitorization of Google Scholar data. Because of this, few studies have empirically documented this phenomenon. This study analyses a large decrease in coverage of documents in the field of Astronomy and Astrophysics that took place in 2019 and its subsequent recovery, using longitudinal data from previous analyses and a new dataset extracted in 2020. Documents from most of the larger publishers in the field disappeared from Google Scholar despite continuing to be available on the Web, which suggests an error on Google Scholar's side. Disappeared documents did not reappear until the following index-wide update, many months after the problem was discovered. The slowness with which Google Scholar is currently able to resolve indexing errors is a clear limitation of the platform both for literature search and bibliometric use cases.

**Introduction**

Academic bibliographic databases, and especially those that generate citation graphs, usually implement document inclusion policies that rarely allow records of documents to be removed once they have entered the system. In a bibliographic database that is intended for literature discovery, coverage stability is a desirable property, if we assume that users intuitively expect a system to retrieve the same documents over time given the same query (in addition to new documents that also meet the search criteria). This property is especially critical for some literature search use cases such as those carried out for systematic reviews, where reproducibility of the process is essential (Gusenbauer & Haddaway, 2020; Haddaway & Gusenbauer, 2020). In a citation index, the disappearance of a document would affect the citation counts of all its cited documents, impeding some types of citation analysis.

Google Scholar continues to be widely used for literature discovery, and sometimes as a data source for research evaluations. Some reasons for this are its comprehensive coverage, and that it is free to access. However, unlike other analogous tools, Google Scholar intentionally operates in a way that does not maintain coverage stability (Delgado López-Cózar et al., 2019). Instead, Google Scholar mirrors the general-purpose Google search engine, and subordinates the continued inclusion of documents in its index to their ongoing availability on the Web (as well as their continued abidance to Google Scholar's technical guidelines). Its documentation declares that this approach was chosen to provide a current reflection of the academic web at any given time (Google Scholar, n.d.-b).

As a result of this policy, coverage in Google Scholar not only increases when new documents indexed, but can also decrease when indexed documents become unavailable to Google Scholar's crawlers. To learn whether documents have stopped being available on the Web, Google Scholar carries out two complete recrawls of its index approximately every year (Google Scholar, n.d.-a). If Google Scholar's crawlers are not able to access a document during one of these recrawls, it is removed from the index.

In some cases, documents that have been removed are still findable in Google Scholar, because they are available from other sources on the Web which Google Scholar also indexes, or because they are cited in other works indexed in the system (in these cases, the document is kept as a *[CITATION]-type* record). However, in other cases, and usually after one of these major index recrawls, documents stop being findable in Google Scholar altogether, with the added consequence that citation counts of the documents that they cite also suffer a decrease (provided that cited reference metadata was available for the removed documents).

If we consider Google Scholar merely as a gateway, a digital *non-place* (Augé, 1995) through which users navigate to the places where academic documents can be accessed, rather than as a data source in its own right that could serve as a record of the inherently cumulative academic knowledge and the interactions that occur between academics, then the decision to not display information about documents that are no longer accessible is understandable (if an airport loses a flight route, it does not make sense to for it to keep displaying information about it). Furthermore, under this point of view, decreasing citation counts may not be an overly concerning issue, as their purpose would only be to serve as one of the parameters that is used to rank documents in a search, a purpose for which a certain amount of inaccuracies in citation counts can probably be tollerated.

However, Google Scholar is perceived as a bibliographic data source by many users, which makes decreasing coverage problematic. For example, researchers sometimes report author-level indicators calculated by Google Scholar in research evaluation processes (hiring, promotion, grant applications…), and seeing these figures decrease overnight and without explanation can be a cause of concern and confusion to them. Probably it is the number of questions about this phenomenon that has led Google Scholar to include a clarification in its help pages about why this occurs (Figure 1).

> ▼ **My citation counts have gone down. Help!**
>
> Google Scholar generally reflects the state of the web as it is **currently** visible to our search robots and to the majority of users. When you're searching for relevant papers to read, you wouldn't want it any other way!
>
> If your citation counts have gone down, chances are that either your paper or papers that cite it have either disappeared from the web entirely, or have become unavailable to our search robots, or, perhaps, have been reformatted in a way that made it difficult for our automated software to identify their bibliographic data and references. If you wish to correct this, you'll need to identify the specific documents with indexing problems and ask your publisher to fix them. Please refer to the technical guidelines.

*Figure 1: Extract from Google Scholar help page that explains why citation counts of documents sometimes decrease in this platform*

The issue of decreasing coverage in Google Scholar is compounded by the fact that there is no public information on the sources that are covered by this search engine. Therefore, users have no way of knowing when certain collections of documents are removed from the index, opening the possibility to situations in which users may decide to rely on this platform based on assumptions regarding its coverage that no longer hold true.

This lack of transparency means that the only way to directly observe coverage loss is through regular monitorization of Google Scholar data: recording states of the index at regular intervals. This makes this phenomenon difficult to analyse, as extracting data from Google Scholar is very time-consuming (Else, 2018), and it is difficult to anticipate which documents are going to be dropped by Google Scholar and when. Because of this, few studies have empirically documented this issue.

In November 2017, while carrying out an annual update of a directory of Spanish academic journals covered by Google Scholar Metrics (an annual product released by Google Scholar that calculates a 5-year h-index for journals), Delgado López-Cózar & Martín-Martín (2018) noticed that the number of available Spanish journals in this product had sharply decreased compared to previous years, breaking the general growing trend observed since 2012: while the 2016 edition of the directory contained 1,101 Spanish journals, in the 2017 edition only 599 journals could be found (Delgado López-Cózar & Martín-Martín, 2019). Journals from all fields disappeared, but Law journals were particularly affected, going from 156 journals in 2016 to 35 journals in 2017. Because Spanish Law journals do not yet have a strong presence on the Web, we turned our attention to the largest bibliographic database that focuses on academic content published in Spain: Dialnet. This database is still the only window through which many journals published in Spain are visible on the Web, and therefore was considered likely to be involved in this *blackout* of Spanish scientific production in Google Scholar. After searching content available from Dialnet in Google Scholar and comparing the results to previous web domain analyses that we had carried in the past (Delgado López-Cózar et al., 2019), it was confirmed that most of the content from Dialnet had disappeared at some point from the search engine. This analysis was published in the Spanish LIS-focused mailing list *Iwetel,* where Dialnet's Technical Director confirmed that they had been aware of this issue since June 2017. Apparently Google Scholar had detected that the metadata of a small batch of old records from Dialnet were inconsistent with metadata for the same documents found in other web sources, and decided to remove most of Dialnet from its index under suspicion of providing incorrect metadata. Dialnet promply fixed this issue, but its records did not return to Google Scholar results until the following complete recrawl of the index was made public in January 2018. Thus, users interested in this content and who knew it to be covered in the past were unknowingly underserviced by Google Scholar for more than half a year. In this case, the issue was more difficult to detect because Dialnet did not contain cited reference metadata that Google Scholar could access and use in its citation graph, and therefore citation counts were not affected. This episode revealed that despite its known errors and limitations (Orduna-Malea et al., 2017), Google Scholar subjects its data to certain quality-control measures. This is, to our knowledge, the first empirically documented case of a large coverage fluctuation in Google Scholar.

In 2019, the editor of the journal *Astronomy & Astrophysics* denounced an apparently similar case (Forveille, 2019). In March 2019 the journal was notified by several researchers that citation counts to documents of this journal had decreased "by an order of magnitude" in their personal Google Scholar profiles. The editor contacted Google Scholar, who acknowledged the error and promised to remedy it. However, this would not be visible in the platform until the next complete recrawl of the index. This resulted in a sharp decrease of the h-index of this journal in the 2019 edition of Google Scholar Metrics, which was computed after Google Scholar was made aware of the issue, but using the yet uncorrected index: while the 2018 edition displayed an h5-index of 115 for this journal, in the 2019 edition this figure dropped to 52. Given the inherent resistance of a high h-index to small random changes in the underlying citation counts (probably the reason why Google Scholar favors this indicator), this signaled a very significant drop in coverage of documents in this field.

The large drop in Astronomy & Astrophysics documents in Google Scholar was also noticed by Martín-Martín et al. (2018, 2021). Analysing a collection of citations to a sample of highly-cited documents from all subject areas collected in 2018 to compare relative differences in coverage in Google Scholar, Scopus, and Web of Science, they found that Google Scholar was able to find 98% of the citations found by Web of Science, and 97% of the citations found by Scopus. Additionally, 30% of all citations were only found in Google Scholar. In 2019 the data was collected again using the same sample of highly-cited documents, but the citations to Astronomy & Astrophysics documents obtained from Google Scholar had radically changed (unlike in other subject categories, where relative differences among data sources remained mostly the same as in 2018): Google Scholar was only able to find 60% of all Web of Science citations, and 60% of all Scopus citations. Since the citation data extracted from Web of Science and Scopus in 2019 contained the same citations that were extracted in 2018, plus the new citations included in these systems between the two points of extraction, this large difference also signaled a significant drop in coverage in Google Scholar.

The datasets extracted from Google Scholar for Martín-Martín et al. (2018, 2021) provide us with an opportunity to analyse this case in more detail. Therefore, the goal of this study is to document this case, to try to find out the cause of this sudden drop in coverage to documents in the field of Astronomy & Astrophysics, and to check whether this issue was resolved in subsequent recrawls of Google Scholar's index.

**Methods**

The datasets extracted from Google Scholar and analysed in Martín-Martín et al. (2018, 2021) were used. These datasets contain the lists of citing documents to a sample of 2515 highly-cited documents from 2006 that Google Scholar released in 2017 with the name *Google Scholar Classic Papers* (GSCP). In this product[1], Google Scholar displayed the top 10 most cited documents published in 2006 in each of 252 subject categories. For more information about this product, see Orduna-Malea et al. (2018).

Since in this study we are particularly interested in the coverage of Astronomy & Astrophysics documents in Google Scholar, only the 10 highly-cited documents in this field in GSCP (Table 1), and the list of citing documents in Google Scholar for each of these documents, are analysed here.

---

1  https://scholar.google.com/citations?view_op=list_classic_articles&hl=en&by=2006

**Table 1: Highly-cited documents in Astronomy & Astrophysics in Google Scholar Classic Papers**

| Art. # | Journal | Vol(issue), pages | DOI |
|---|---|---|---|
| 1 | The Astronomical Journal | 131(2), 1163-1183 | 10.1086/498708 |
| 2 | Monthly Notices of the Royal Astronomical Society | 365(1) 11-28 | 10.1111/j.1365-2966.2005.09675.x |
| 3 | Astronomy & Astrophysics | 447(1), 31-48 | 10.1051/0004-6361:20054185 |
| 4 | Monthly Notices of the Royal Astronomical Society | 370(2) 645-655 | 10.1111/j.1365-2966.2006.10519.x |
| 5 | Nuclear Physics A | 777, 1-4 | 10.1016/j.nuclphysa.2005.06.010 |
| 6 | The Astrophysical Journal | 651(1), 142-154 | 10.1086/506610 |
| 7 | Physical Review D | 74(12), 123507 | 10.1103/PhysRevD.74.123507 |
| 8 | The Astronomical Journal | 131(4), 2332-2359 | 10.1086/500975 |
| 9 | Monthly Notices of the Royal Astronomical Society | 368(1) 2-20 | 10.1111/j.1365-2966.2006.10145.x |
| 10 | Monthly Notices of the Royal Astronomical Society | 372(3) 961-976 | 10.1111/j.1365-2966.2006.10859.x |

The list of documents that cite each document in Table 1 was extracted from Google Scholar in three different occasions: April-May of 2018, May-June 2019, and April 2020. To do this, a custom scraper was used (Martín-Martín, 2018).

The metadata extracted from Google Scholar for each of these citing documents was enriched by complementing it with metadata available in the HTML meta tags of the webpages where Google Scholar found these documents, and the metadata available in CrossRef's and arXiv's public APIs. A DOI was found for 79% of the citations in the 2018 dataset, 76% of the citations in the 2019 dataset, and 77% of the citations in the 2020 dataset.

Citations across the three datasets were matched based on (in this order) Google Scholar's internal document identification codes, the URLs of the webpages were Google Scholar found the citing documents, the DOIs of the citing documents, and a combination of title and author similarity, in a similar way as described in Martín-Martín et al. (2018, 2021):
1. For each pair of datasets A and B and a seed highly-cited document X, all citing documents with a document id (Google Scholar ID, URL, and DOI) that cite X according to A were matched to all citing documents with a corresponding document id that cite X according to B.
2. For each of the unmatched documents citing X in A and B, a further comparison was carried out. The title of each unmatched document citing X in A was compared to the titles of all the unmatched documents citing X in B, using the restricted Damerau-Levenshtein distance (optimal string alignment) (Damerau, 1964; Levenshtein, 1966). The pair of citing documents which returned the highest title similarity (1 is perfect similarity) was selected as a potential match. This match was considered successful if either of the following conservative heuristics was met:
    1. The title similarity was at least 0.8, and the title of the citing document was at least 30 characters long (to avoid matches between short, undescriptive titles such as "Introduction").
    2. The title similarity was at least 0.7, and the first author of the citing document was the same in A and B.

First, citations in the 2019 dataset were matched to citations in the 2018 dataset. The result of that matching was in turn matched to the citations in the 2020 dataset.

**Results**

A simple observation of the citation counts reported by Google Scholar over the years for the 10 highly-cited documents in the field of Astronomy & Astrophysics in GSCP already reveals a large fluctuation (Figure 2).

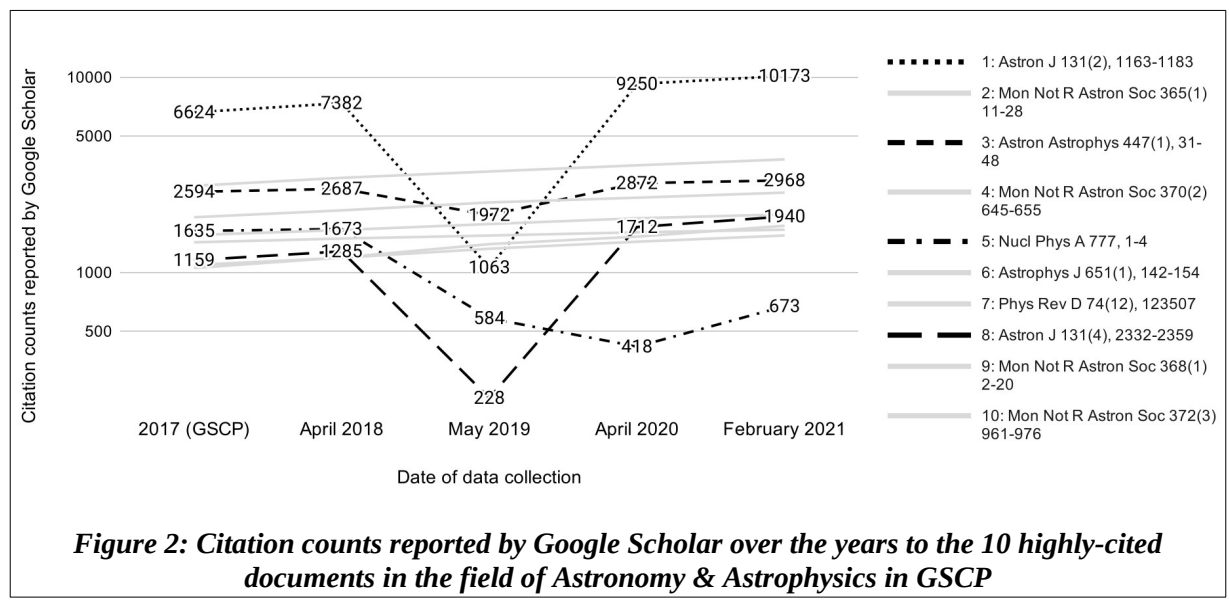

*Figure 2: Citation counts reported by Google Scholar over the years to the 10 highly-cited documents in the field of Astronomy & Astrophysics in GSCP*

Four of the 10 documents suffered a sharp decresase in citations in 2019. In the most extreme case (document #1), the citation count in 2019 had decreased by 6,319 citations respect to the count reported in 2018. It is important to note that this document was published not in the journal *Astronomy & Astrophysics*, but in *The Astronomical Journal*, published by IOP Science. Indeed, of the four documents that show a drop in citation counts in 2019, only one was published in *Astronomy & Astrophysics* (document #3). There is a second article from *The Astronomical Journal* with decreased citation counts in 2019 (document #8), and another one published *Nuclear Physics A* (document #5). This shows that *Astronomy & Astrophysics* was not the only journal to be affected by the drop in coverage in 2019, which is expected, since whichever documents disappeared from Google Scholar likely cited articles from various journals in the field.

Six of the ten documents (#2, #4, #6, #7, #9, #10), however, do not show clear signs of a coverage drop: their citation counts grew each year. Four of these documents were published in *Monthly Notices of the Royal Astronomical Society* (Oxford University Press), one in The Astrophysical Journal (IOP Science), and one in Physical Review D (American Physical Society). This does not rule out the possibility that they lost citations in 2019, only that the growth/loss balance was positive. Nevertheless, this suggests that some documents could have been more affected than others.

Of the four documents that clearly lost citations in 2019, three of them seem to recover them by 2020 and continue receiving citations in 2021. This suggests that the coverage loss was indeed temporary, and that citations were recovered at some point between summer of 2019 and spring of 2020, probably after the second complete recrawl of the index in 2019. One of the documents (#5), however, does not recover the amount of citation counts that were reported in 2017 and 2018, even by 2021. After closer examination of this document, titled "The solar chemical composition" and published in *Nuclear Physics A*, it was discovered that there are actually two documents with the same title and the same authors, published in two different journals (Figure 3). We believe it is probable that in 2017 and 2018, these two documents were incorrectly merged into one record (combining the citations of the two), and that in 2019 they were separated, as they remain in 2021. Since this is a different kind of Google Scholar error than the one we are documenting here, this document and its citations were excluded from further analysis in this study.

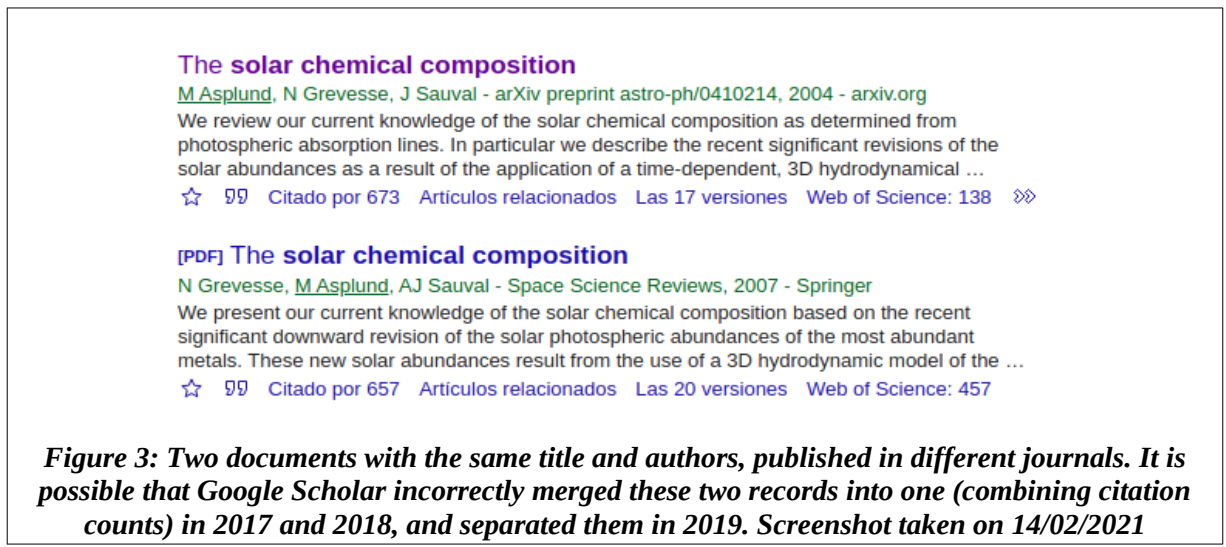

*Figure 3: Two documents with the same title and authors, published in different journals. It is possible that Google Scholar incorrectly merged these two records into one (combining citation counts) in 2017 and 2018, and separated them in 2019. Screenshot taken on 14/02/2021*

We analysed the list of documents that cited each of the nine highly-cited document according to Google Scholar at three points in time: 2018, 2019, and 2020. In 2018, 21,907 citations were extracted. In 2019, 15,042 were extracted, while in 2020, 25,195 citations were found in Google Scholars to these nine documents. Of the 21,907 citations found in 2018, 8,840 (40%) were missing in 2019. In 2020, however, 96% of the citations available in the 2018 dataset had reappeared.

To find out exactly which documents caused the decrease in citation counts, the citations found in 2018 to each of the nine highly-cited documents were grouped by the publisher of the citing document, and by whether or not the citation was also found in 2019 and 2020 (Figure 4). Missing documents are mostly concentrated in document #1, #8, and #3, while the other six documents are affected to a much lower extent.

Missing documents in 2019 are discributed across many of the largest publishers in the field (Figure 4). They are also distributed across all publication years. Therefore, because we know that these large publishers did not actually disappear from the Web in 2019, and it is unlikely that they concertedly modified the metadata of their documents or their crawler access rules so as to not comply with Google Scholar's technical guidelines, it is likely that the error originated at Google Scholar's side. The exact cause of the error is, nevertheless, unknown.

There are several possibilities as to why some citations from 2018 are still missing in the 2020 dataset: some may correspond to documents that did actually disappear from the web (particularly those in the category "Other", in which some of the documents are hosted in less stable websites). Some others may correspond to duplicate records found in 2018 that were subsequently merged.

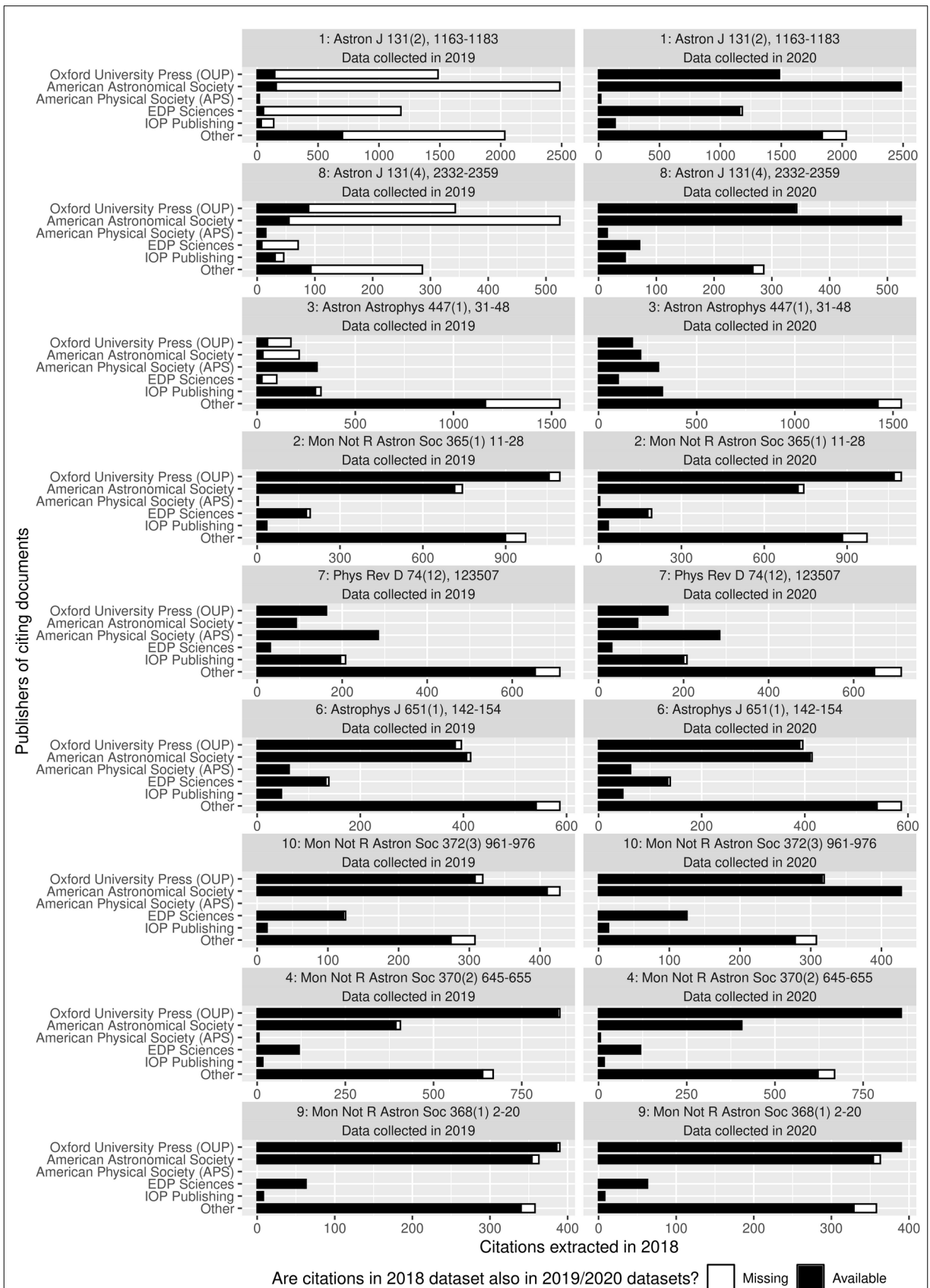

*Figure 4: Citations found in 2018 for each of the 9 highly-cited documents, grouped by publisher of citing document and by whether or not the citation was found in 2019 and 2020*

The citing documents that were present in the three datasets (data extracted in 2018, 2019, and 2020) came with citation counts of their own, which provides us with an opportunity to gauge to what extent each publisher was affected by the loss in coverage in 2019 using a larger sample than the 10 highly-cited documents in GSCP. In this regard, documents published by EDP Sciences (publisher of the journal *Astronomy & Astrophysics*) were severely affected by the loss of coverage (Figure 5), followed by documents published by the American Astronomical Society, which were affected to a much lower extent. Out of the 724 documents published by EDP Sciences that were available in the three datasets, 404 of them (58%) reported at least 10 citations less in the 2019 dataset than in the 2018 dataset, whereas out of the 2,604 documents published by the American Astronomical Society present in the three datasets, 141 (5%) reported at least 10 citation less in the 2019 dataset than in the 2018 dataset. In the rest of publishers, lower citations in the 2019 dataset than in the 2018 dataset were even more uncommon.

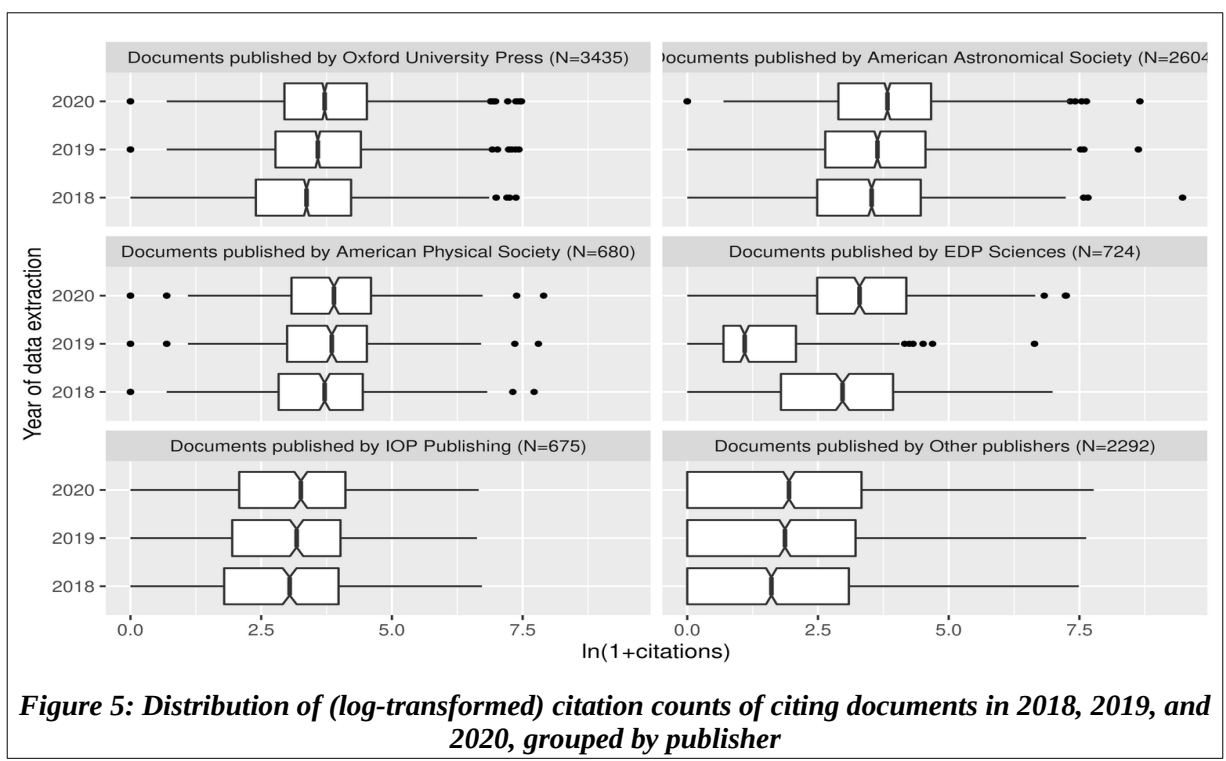

*Figure 5: Distribution of (log-transformed) citation counts of citing documents in 2018, 2019, and 2020, grouped by publisher*

**Discussion and conclusions**

The results confirm that this is another case of a large coverage fluctuation in Google Scholar. Similarly to what happened during the Dialnet blackout event, users interested in Astronomy and Astrophysics content during 2019 who expected Google Scholar to have a comprehensive coverage of this field could have been unknowingly underserviced for a period of 6 to 9 months. Unlike in the Dialnet event, however, the effects were quickly felt because of the drastic drop in citation counts in documents of the area, especially in the journal *Astronomy & Astrophysics*, which are confirmed in this study. Furthermore, although this analysis has not been able to discern the specific cause of the error, the fact that documents from many large publishers disappeared from the platform despite continuing to be available on the Web points to a mistake on Google Scholar's side.

Although there are reports that Google Scholar sometimes contact content providers when an issue like this arises (Delgado López-Cózar & Martín-Martín, 2018), the slowness with which it is currently able to resolve this kind of issues is a clear limitation of the platform for literature search use cases as well as for bibliometric use cases.